\documentclass[preprint2,10.5pt]{aastex}
\usepackage{url}
\begin{document}
\title{\large{\rm{THE IMPACT OF CONTAMINATED RR LYRAE/GLOBULAR CLUSTER PHOTOMETRY ON THE DISTANCE SCALE}}}
\author{D. Majaess$^1$, D. G. Turner$^1$, W. Gieren$^2$,D. J. Lane$^1$}
\affil{$^1$Saint Mary's University, Halifax, NS, Canada.}
\affil{$^2$Universidad de Concepci\'on, Concepci\'on, Chile.}
\email{dmajaess@cygnus.smu.ca}

\begin{abstract} 
RR Lyrae variables and the stellar constituents of globular clusters are employed to establish the cosmic distance scale and age of the universe.  However, photometry for RR Lyrae variables in the globular clusters M3, M15, M54, M92, NGC2419, and NGC6441 exhibit a dependence on the clustercentric distance.  For example, variables and stars positioned near the crowded high-surface brightness cores of the clusters may suffer from photometric contamination, which invariably affects a suite of inferred parameters (e.g., distance, color excess, absolute magnitude, etc.).  The impetus for this study is to mitigate the propagation of systematic uncertainties by increasing awareness of the pernicious impact of contaminated and radial-dependent photometry.
 \end{abstract}


\section{{\rm \footnotesize INTRODUCTION}}
\begin{figure}[!t]
\begin{center}
\includegraphics[width=5.5cm]{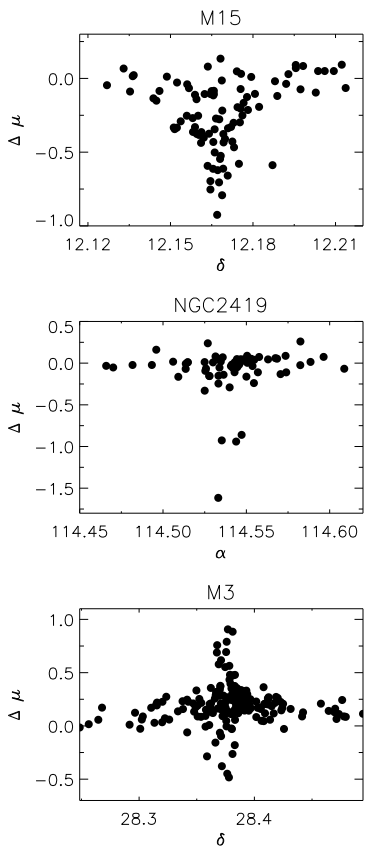} 
\caption{Relative distance moduli for RR Lyrae variables in M15, M3, and NGC2419 exhibit a positional dependence.  The effect of photometric contamination becomes pronounced with decreasing clustercentric distance.  The relative distance moduli were computed according to Eqn.~(\ref{eqn-wvi}). }
\label{fig-dist}
\end{center}
\end{figure}

Extragalactic photometry for classical Cepheids in M33, M81, M101, and M106 implies that at a fixed pulsation period the variables brighten ($W_{VI_c}$) with decreasing galactocentric distance \citep[e.g.,][]{ke98}.  A debate persists as to whether the trend is tied to a metallicity gradient, or arises from photometric contamination \citep[][and references therein]{ma10,ma09,ma11b,ku12}.  The degeneracy exists because the surface brightness, density, and metal abundance \citep[][their Fig.~3]{an04} all increase with decreasing galactocentric distance.  A similar behavior is observed in photometric data for RR Lyrae variables in the globular cluster M15 \citep{ma11}, whereby distance moduli established for the variables likewise exhibit a radial dependence (Fig.~\ref{fig-dist}).  However, no linear metallicity gradient exists across the globular cluster, and thus the degeneracy plaguing the extragalactic Cepheids is absent. Consequently, observations of variables near the core of M15 suffer from photometric contamination \citep[see also][]{ya94,st05}.  A dependence of the photometry on the clustercentric distance may complicate conclusions drawn from Bailey (period-amplitude) diagrams for globular clusters \citep[][and see also \citealt{ku11} who describe the impact of the Blazhko effect on that diagram]{co08,sme11}.

In this study, photometry for RR Lyrae variables in the globular clusters M3, M54, M92, NGC2419, NGC6441, and M15 are shown to exhibit a dependence on clustercentric distance (Figs.~\ref{fig-dist},~\ref{fig-r}).  It follows that the color-magnitude diagrams are compromised (Fig.~\ref{fig-cmds}).  Clustercentric photometry has a direct impact on the establishment of Galactic/extragalactic distances (e.g., brightness of the Horizontal Branch), the Hubble constant ($H_0$), the age of the universe, and cosmological models \citep[e.g., $\sigma_w\sim 2\times H_0$,][]{mr09}.

\section{{\rm \footnotesize ANALYSIS}}
\subsection{{\rm \footnotesize RR LYRAE VARIABLES}}
Distance moduli for RR Lyrae variables may be inferred from a reddening-free Wesenheit function ($W_{VI_c}$).  \citet{so03} cite the following $VI_c$ Wesenheit function for LMC RR Lyrae (ab) variables:
\begin{eqnarray}
\nonumber
W_{VI_c}=(-2.75\pm0.04)\log{P}+(17.217\pm0.008) 
\end{eqnarray}
The empirical slope of the function is consistent with that predicted by models \citep{di04}.   The distance modulus for each RR Lyrae variable was evaluated via:
\begin{eqnarray}
\label{eqn-wvi}
\nonumber
m-M= 5\log{d}-5 \\
\nonumber
W_{VI_c}-W_{VI_c,0}= 5\log{d}-5 \\
\nonumber
V-R_{VI_c}(V-I_c)-\alpha \log{P_0}-\beta=\mu_0 \\
V-2.55(V-I_c)+2.73 \log{P_0}-\beta=\mu_0
\end{eqnarray}
$R_{VI_c}=2.55$ is the ratio of total to selective extinction \citep[][and references therein]{di04,fo07}, $\alpha$ is the slope of the Wesenheit function, and $\beta$ is chosen to be arbitrary (relative distance).   A mean of two estimates for the Wesenheit slope (LMC/IC4499) was adopted ($\alpha_{VI_c}=-2.73\pm0.03$).  Only RR Lyrae variables sampled beyond the core of IC4499 were analyzed \citep{wn96}. 

Relative distances were evaluated for RR Lyrae variables discovered in M15 \citep{co08}, M3 \citep{be06}, NGC2419 \citep{di11}, M92 \citep{ko01}, M54 \citep{ls00}, and NGC6441 \citep{pr03}.   The periods of RR Lyrae (c) variables pulsating in the first overtone were fundamentalized via $\log{P_0}\sim\log{P}+0.13$, where $P_0$ is the fundamental mode period.  RA/DEC-distance diagrams were constructed for variables in M15, M3, and NGC2419 (Fig.~\ref{fig-dist}).  As expected the distances become particularly discrepant as the cluster core is approached. The stellar density and surface brightness increase markedly as a function of decreasing clustercentric distance.  For M15 and NGC2419 a similar trend is observed, namely that distances for stars near the core appear underestimated.  A significant fraction of the variables in M15 are contaminated, and numerous stars are likewise affected in M3.  Four stars in NGC2419 appear particularly compromised (i.e., $\Delta \mu \la -0.5$), and less-pronounced contamination persists to near the periphery.  NGC2419 exhibits a low stellar density\footnote{Globular Cluster Database, W. E. Harris, McMaster U.: \url{http://physwww.mcmaster.ca/~harris/mwgc.dat}} relative to the other globular clusters analyzed, however, the density of stars per cubic arcminute is large since the cluster is distant ($d\sim83$ kpc, \S \ref{s-ngc2419}). 

\begin{figure*}[!t]
\begin{center}
\includegraphics[width=16cm]{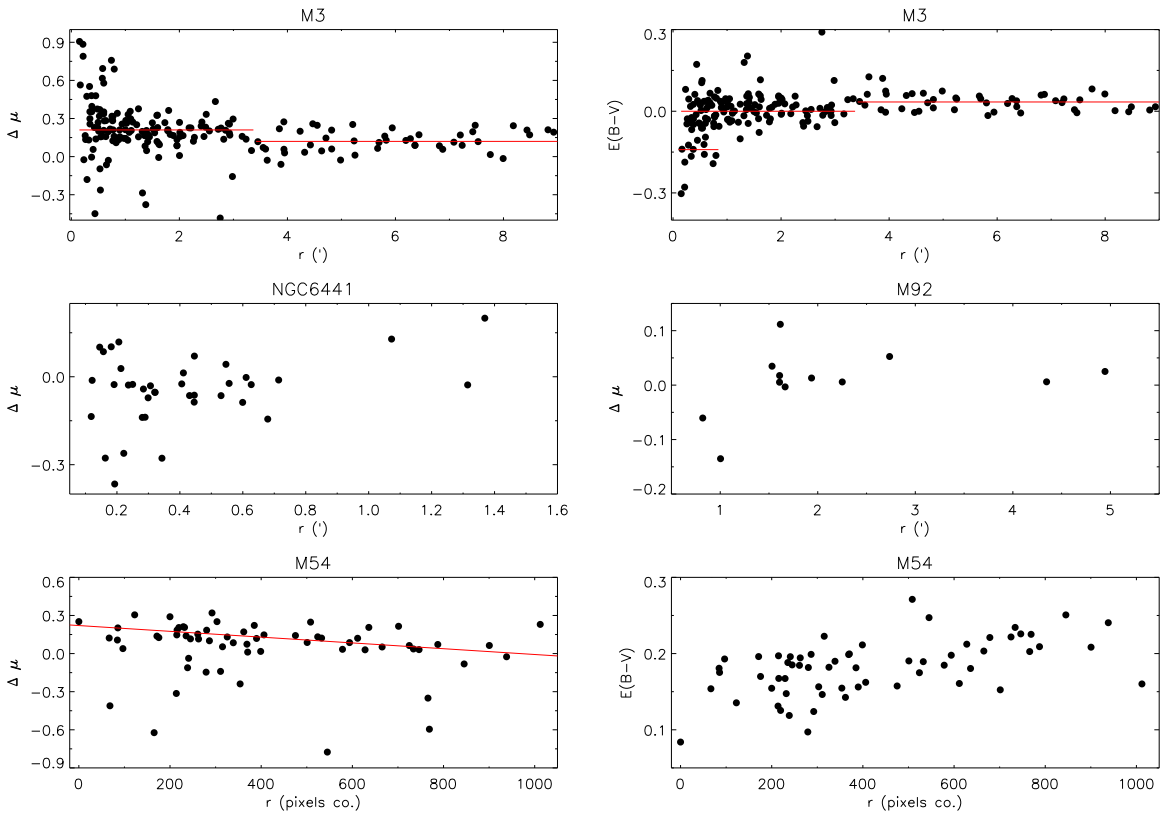} 
\caption{Relative distance moduli and reddenings for RR Lyrae variables in M3, NGC6441, M92, and M54 exhibit a dependence on clustercentric distance. Photometric contamination appears most pronounced near $r\sim0 \arcmin$.  The relative distance moduli were computed according to Eqn.~(\ref{eqn-wvi}). }
\label{fig-r}
\end{center}
\end{figure*}

Fig.~\ref{fig-r} features parameters for RR Lyrae variables in M3, M92, M54, and NGC6441 plotted as a function of clustercentric distance.  RR Lyrae variables in NGC6441 and M92 display an analogous trend to that noted for NGC2419 and M15.  Stars near the core exhibit smaller distances than objects occupying the cluster periphery.  Seven stars within $r\la 0.5 \arcmin$ (NGC6441) appear particularly contaminated (Fig.~\ref{fig-r}).  NGC6441 is projected along a line of sight where the field density is high ($\ell,b\sim354,-5 \degr$).  RR Lyrae variables in M54 display a similar trend to those in M3, whereby the moduli and reddenings computed appear nearer and larger with increasing clustercentric distance, respectively.  It is unclear whether effects associated with the edge of CCD detectors play a role. Thus photometric contamination is not the only source of concern that imposes a clustercentric dependence (further research is necessary).  Stars in M3 located $r\ga3.5\arcmin$ appear closer than those near the core (Fig.~\ref{fig-r}).  A period-reddening relation indicates the variables are redder than objects located near the core, which are anomalously blue (Fig.~\ref{fig-r}).  The period-color relation does not account for the width of the instability strip, thus a symmetric distribution consisting of negative reddenings is expected for a population exhibiting $E(B-V)\sim0$.  However, deviations from symmetry are observed for stars $r\la1\arcmin$ (Fig.~\ref{fig-r}). 

Table~\ref{table1} highlights the difference between the distance computed using all RR Lyrae variables from those located near the periphery ($\bar{\mu}-\mu_{r}$).  The offset is largest for M15 ($\bar{\mu}-\mu_{r}\sim-0.25$) and smallest for M92.     M3 exhibits a positive offset, and the average value for the five clusters featuring a negative offset is $\bar{\mu}-\mu_{r}\sim-0.1$.  Admittedly, the differences evaluated are subjective and uncertain owing partly to the neglect of superposed trends (M54, Fig.~\ref{fig-r}), and an inability to overcome poor statistics and assess the extent of the compromised photometry (NGC6441, M92, Fig.~\ref{fig-r}).  Nevertheless, the results imply a potential $\sim10\%$ systematic uncertainty in age.  The clustercentric dependence therefore hinders efforts to determine reliable parameters for RR Lyrae variables, and age estimates for the host clusters.  The clusters and their RR Lyrae variable constituents are likewise employed to establish the distance scale, from which estimates of $H_0$ and the age of the universe stem.   Globular clusters themselves set a crucial lower limit to the age of the universe.  All potential biases (regardless of their relative impact) must be assessed in the pursuit of precision cosmology.

\subsection{{\rm \footnotesize GLOBULAR CLUSTER CMDS}}
\begin{figure*}[!t]
\begin{center}
\includegraphics[width=15cm]{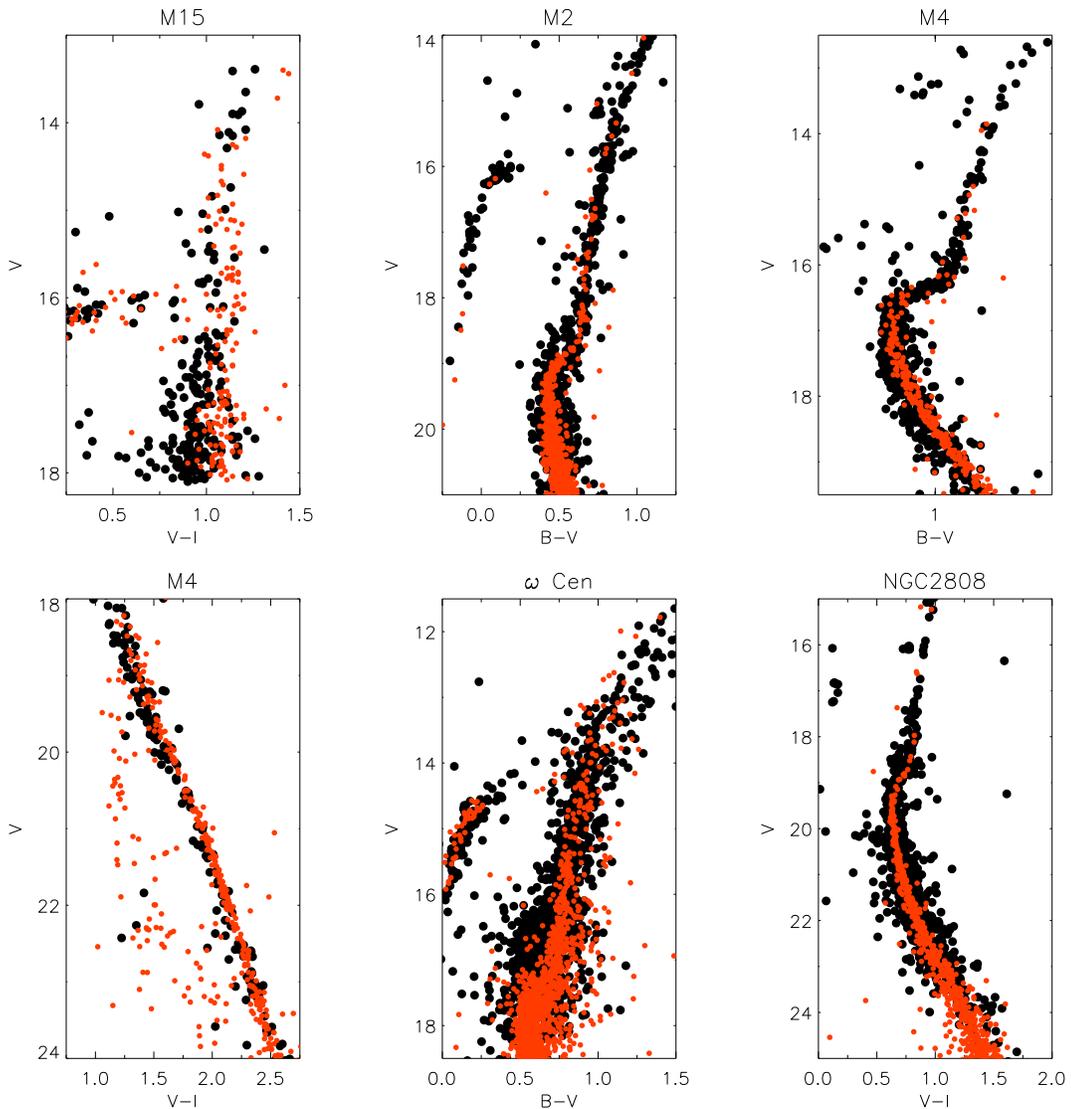} 
\caption{Color-magnitude diagrams for M15, M2, M4, $\omega$ Cen, and NGC2808.  Black dots define stellar constituents closest to the cluster core, whereas red dots characterize a subset of stars beyond the core.  The morphology displayed in the color-magnitude diagrams is sensitive to the clustercentric distance \citep[see also discussion in][]{ya94,st05}.}
\label{fig-cmds}
\end{center}
\end{figure*}

Color-magnitude diagrams for globular clusters are used to determine their distance (e.g., horizontal branch, subdwarfs) and age (isochrone fitting).  The radial dependencies noted for the RR Lyrae variables (Figs.~\ref{fig-dist},~\ref{fig-r}) promulgate into the color-magnitude diagrams.  Color-magnitude diagrams are tabulated for M15 \citep{ya94}, M2 \citep{lc99}, M4 \citep{ib99,mo02}, $\omega$ Cen \citep{re04}, and NGC2808 (Fig.~\ref{fig-cmds}).  Groups of stars sampling near the core and periphery of the clusters are overplotted.  The severity of the contamination depends on the distance and stellar density (stars per cubic arcminute), data-processing software, and contribution to the seeing owing to the instrumentation (CCD, telescope, wavelength) and environment (e.g., ground or space-based observations), etc. \citep[see also][]{ya94,st05}.  PSF photometry does not correct for multiple stars coincident well within the FWHM of the system.  The clustercentric effects  (Fig.~\ref{fig-dist},~\ref{fig-r}) impose a degeneracy which complicates the analysis of multiple populations and potential radial trends which are population-dependent (e.g., $\omega$ Cen), although for the latter mixing over such a large timescale and mass segregation likewise require consideration.  The signatures of multiple populations may be mimicked by photometric contamination. 

Photometry acquired for M13 from the Abbey Ridge Observatory (ARO\footnote{The ARO is a 0.3-m robotic optical facility used to detect supernovae and conduct variable star research \citep[e.g.,][]{lg05,la08}.}) confirms the trends observed in Fig.~\ref{fig-cmds}.  For example, as expected photometry near the core is artificially brighter and less complete owing to the increased sky brightness.  The ARO's comparatively small aperture telescope, in tandem with typically large local seeing, conspire to create a sizable FWHM ($\sim2.5-3.0 \arcsec$).  HST seeing is more than an order of magnitude smaller, and for that reason the ARO data were acquired to provide an empirical means to evaluate contamination effects directly.  The trends observed are not unique to a given dataset cited, but may be present (to an extent) in most globular cluster photometry.

\begin{figure*}[!t]
\begin{center}
\includegraphics[width=7.8cm]{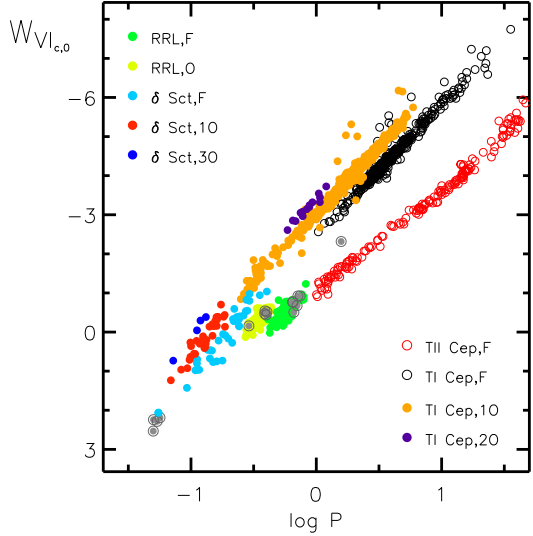} 
\caption{The universal Wesenheit template leverages the statistical weight of the entire variable star demographic to constrain distances and pulsation modes for target populations (e.g., NGC2419).  SX Phe, RR Lyrae, and Type II Cepheid variables in NGC2419 which lie $r\ga2.5\arcmin$ are shown (gray dots).}
\label{fig-w}
\end{center}
\end{figure*}

\subsection{{\rm \footnotesize DISTANCE TO NGC2419}}
\label{s-ngc2419}
A distance to NGC2419 may be derived which is less sensitive to the trends highlighted in Fig.~\ref{fig-dist}.  Absolute distances may be secured by anchoring the \citet{so03} RR Lyrae $VI_c$ Wesenheit function to the LMC distance established by \citet[][$\mu_{0,LMC}=18.43\pm0.03$]{ma11}\footnote{A mean computed from $\sim5\times 10^2$ LMC distance estimates tabulated in the NASA/IPAC Extragalactic Database of galaxy Distances (NED-D) supports that result \citep[][see also Fig.~2 in \citealt{fm10}]{sm11}.}:
\begin{eqnarray}
\nonumber
W_{VI_c} (RRL,LMC)=(-2.75\pm0.04)\log{P} \\ \nonumber +(17.217\pm0.008)  \\
\nonumber
W_{VI_c,0}= \alpha \log{P_0}+\beta \\
\nonumber
W_{VI_c}-W_{VI_c,0}= 5\log{d}-5 \\
\nonumber
(-2.75\pm0.04)\log{P} +  (17.217\pm0.008) - \\ \nonumber  (\alpha \log{P_0}+\beta)=\mu_0 
\end{eqnarray}
$\beta$ can be solved by re-arranging the aforementioned equation.  The absolute Wesenheit function characterizing RR Lyrae variables is thus:
\begin{eqnarray}
\label{eqn-wvi2}
\nonumber
W_{VI_c,0}\simeq(-2.73\pm0.03)\log{P_0}-(1.21\pm0.03)
\end{eqnarray}
\citet{ma11} derived the LMC distance via a universal Wesenheit template, which leverages the statistical weight of the entire variable star demographic to establish precise ($<5$\%) distances.  The template capitalizes upon HST \citep{be07}, VLBA, and HIP \citep{vl07} geometric distances for SX Phe, $\delta$ Scuti, RR Lyrae, Type II and classical Cepheid variables.  Specifically, the template is anchored to 8 SX Phe and $\delta$ Sct variables \citep[HIP,][]{vl07}, 4 RR Lyrae variables \citep[HIP and HST,][]{be02,vl07}, 2 Type II Cepheids \citep[HIP,][]{vl07}, and 10 classical Cepheids \citep[HST,][]{be07}.   The template may be refined by employing new HST parallaxes for RZ Cep, XZ Cyg, SU Dra, UV Oct, VY Pyx, and $\kappa$ Pav \citep{be11}.  However, certain of the aforementioned stars lack reliable mean $VI_c$ photometry, which is necessary for tabulating Wesenheit magnitudes.  The desired observations are presently being acquired via facilities operated by the AAVSO (T. Krajci \& A. Henden/BSM).  Type II Cepheids detected in the less-crowded outer region of M106 \citep{ma06,ma09} were likewise incorporated into the Wesenheit template owing to the availability of a precise maser distance \citep[][and references therein]{ma06}.  The Wesenheit approach obviates the need for uncertain reddening corrections, and thus the maser distance is used directly. Admittedly, the approach supersedes one of the authors (DM) prior first-order (yet ultimately incorrect) effort to unify variables of the instability strip in order to establish reliable distances \citep[][and references therein]{ma11}.

Equation~\ref{eqn-wvi2} may be employed to establish the distance to NGC2419, however, a preferred approach is to simultaneously use the SX Phe, RR Lyrae, and Type II Cepheid variables discovered by \citet{di11}.  An advantage of the universal Wesenheit template is that known pulsation modes for RR Lyrae (ab) constituents may be employed to constrain that (potentially ambiguous) parameter for the SX Phe variables.  The universal Wesenheit template is shown as Fig.~\ref{fig-w}.  The template features Galactic \citep[][and references therein]{ma11} and M106 data, in addition to LMC variables cataloged by OGLE \citep[e.g.,][]{ud99,so08} which were incorporated into the template using the LMC distance derived by \citet{ma11}. Designations for several variables tabulated in the OGLE survey are inconsistent with their Wesenheit magnitudes (e.g., overtone classical Cepheids on the fundamental mode ridge, etc.), yet these objects constitute the minority (see also the break between $\delta$ Scuti and overtone classical Cepheids). 

To mitigate the effects of contaminated photometry stars $r\ga2.5\arcmin$ (subjective) from the core of NGC2419 were utilized to deduce the distance.  Stars exhibiting anomalous Wesenheit positions were not included in the determination.  The offset from the Wesenheit template implies $\mu_0\ga19.56\pm0.05$ for NGC2419, which is consistent with estimates tabulated by Harris and \citet{di11}.  The distance is merely evaluated for illustrative purposes since the maximum extent of the contamination is unclear (Fig.~\ref{fig-dist}).  The analysis reveals that the Type II Cepheid (V18) is either contaminated, a binary system, anomalous, or not a cluster member (Fig.~\ref{fig-w}).  The binary hypothesis is supported by the discovery of numerous binary classical and Type II Cepheids\footnote{No \textit{bona fide} binary RR Lyrae variables have been discovered/established in the LMC \citep[][see also \citealt{pi12}]{pr08}.} in the Galaxy and LMC \citep{ev91,ev92,sz03,so08}.  The three SX Phe variables lie on the Wesenheit ridge tied to fundamental mode pulsators (Fig.~\ref{fig-w}).  The stars are coincident with the absolute Wesenheit magnitude for the prototype of the class \citep[SX Phe, see also][]{ma11}.    SX Phe and other metal-poor population II $\delta$ Scutis lie toward the short-period extension of the Wesenheit ridge characterizing population I $\delta$ Scutis. 

\subsection{{\rm \footnotesize CONCLUSION}}
The analysis reaffirms that a (deleterious) clustercentric distance dependence exists in photometry associated with the globular clusters M15, M2, M3, M54, M92, NGC2419, NGC6441, $\omega$ Cen, M4, M13, and NGC2808 (Figs.~\ref{fig-dist},~\ref{fig-r},~\ref{fig-cmds}).  Distances inferred from RR Lyrae variables and cluster stars are aversely affected.  The impact may be reduced by remaining cognisant of the trends exhibited in Figs.~\ref{fig-dist},~\ref{fig-r},~\ref{fig-cmds} \citep[see also][]{ya94,st05}.  The principal impetus of the present work is to increase awareness concerning clustercentric photometry in order to mitigate the propagation of biases into research on RR Lyrae variables, and to serve as a catalyst for additional discussion and continued research.  Photometric contamination, in harmony with the challenges of establishing precise, standardized, multi-epoch, multi-band photometry: are foremost among sources of uncertainty associated with the distance scale \citep[see also][]{ma10,ma11b}.   Admittedly, the radial dependencies in the globular cluster photometry were partly overlooked by one of the authors (DM) previously \citep[][and references therein]{ma11}. 

The conclusions do not mitigate the broader importance of the studies discussed.  For example, the datasets contain sizable numbers of newly discovered variable stars \citep[e.g., SX Phe, RR Lyrae, and Type II Cepheid variables,][]{cl01,sa09}.\footnote{\url{http://www.astro.utoronto.ca/~cclement/read.html}}  The pulsation periods determined place invaluable constraints on models since period-change is a proxy for stellar evolution ($dR/dt$).  Such analyses have hitherto yielded \textit{seminal} results for RR Lyrae and Cepheid variables \citep{lc99,tu06,ku11,ju12,ne12}.

\begin{deluxetable}{lccc}
\tablewidth{0pt}
\tabletypesize{\footnotesize}
\tablecaption{\footnotesize{Relative RR Lyrae Distances}}
\tablehead{\colhead{Cluster}  & \colhead{$\bar{\mu}-\mu_{r}$}}
\startdata
M15	&	-0.25: \\
NGC2419	&	-0.11: \\
M3	&	0.07:	 \\
M54	&	-0.08:	\\
NGC6441	& -0.08:	\\
M92	&	-0.01:		
\enddata
\label{table1}
\tablenotetext{1}{$\bar{\mu}-\mu_{r}$ defines the distance spread between the average computed using all RR Lyrae variables and only those near the periphery.} 
\end{deluxetable}

\subsection*{{\rm \scriptsize ACKNOWLEDGEMENTS}}
\scriptsize{DM is grateful to the following individuals and consortia whose efforts lie at the foundation of the research: OGLE (A. Udalski, I. Soszy{\'n}ski), H. Smith (RR Lyrae Stars), the Padova Globular Cluster Group, CDS, arXiv, and NASA ADS.  WG is grateful for support from the Chilean Center for Astrophysics FONDAP 15010003 and the BASAL Centro de Astrofisica y Tecnologias Afines (CATA) PFB-06/2007.}

\end{document}